\def\G0{G_{\scriptscriptstyle{\!\otimes}}}
\documentstyle[multicol,aps,epsfig]{revtex} 

\tighten       

\begin{document}
\draft

\title{First principles theory of inelastic currents in a 
scanning tunneling microscope}
\author{K. Stokbro$^{1}$, Ben Yu-Kuang Hu$^{1}$, C. Thirstrup$^{2}$ and X. C. Xie$^{3}$}
\address{$^1$Mikroelektronik Centret, Danmarks Tekniske Universitet, 
Bygning 345\o , DK-2800 Lyngby, Denmark.}
\address{$^2$Surface and Interface Laboratory, RIKEN, Saitama 351,
Japan.}
\address{$^3$ Department of Physics, Oklahoma State University,
  Stillwater, OK 74078, U.S.A.}

\date{\today}
\maketitle

\begin{abstract}
A first principles theory of inelastic tunneling between 
a model probe tip and an atom adsorbed on a surface is presented, 
extending the elastic tunneling theory of Tersoff and Hamann.
The inelastic current is proportional to
the change in the local density of states at the center of the tip
due to the addition of the adsorbate.
We use the theory to investigate the vibrational heating 
of an adsorbate below an STM tip.
We calculate the desorption rate of H
from Si(100)-H(2$\times$1) as function of the sample bias and tunnel
current, and find excellent agreement with recent experimental data.
\end{abstract}
\begin{multicols}{2}

\narrowtext
Atom manipulation using scanning tunneling microscopes (STMs) has
been reported for a wide range of systems\cite{StEi91,Av95}. The
manipulation of atoms has been attributed to various origins: direct forces 
between the tip and the surface\cite{EiSc90,LaMeRi97}, indirect 
interaction through the tip induced electric 
field\cite{WhStDrCe91,UcHuGrAo93} and inelastic scattering by electrons
tunneling from the tip to the surface\cite{ShWaAbTuLyAvWa95,StReHoGaPeLu97}.
The last mechanism, for example, is thought to account for 
the reversible transfer of a Xe atom between a Ni(110) surface and a
W tip\cite{EiLuRu91,GaPeLu92,BrHe94}, hydrogen desorption from hydrogen 
passivated silicon\cite{ShWaAbTuLyAvWa95}, and
dissociation of O$_2$ molecules on a Ni(110) surface\cite{StReHoGaPeLu97}.
However, it is often not evident from the experimental data which
mechanism dominates, and input from theoretical calculations is vital 
to establish the microscopic mechanisms behind atomic manipulation on
surfaces.

In this paper we present a first principles method for calculating the
inelastic current in a STM tunnel junction due to inelastic
scattering of the tunneling electrons with an adsorbate induced resonance.  
Our method is based on the Tersoff-Hamann approximation for
the STM tunnel junction\cite{TeHa85} and builds on previous
work on inelastic tunneling in STM\cite{Ga97,GaPeLu97,SaPePa94}.
The main result of this communication, Eq.~(\ref{eq:iinelas}),  is an
expression for the inelastic current  
in terms of the partial local density of states (DOS) of the 
adsorbate wave function at the position of the tip.  
This expressions allows us to calculate the current and bias
dependence of the desorption rate of H from the monohydrate Si(100) surface, 
by solving a Pauli master rate equation for transitions between the 
different vibrational levels of the adsorbate. In a recent letter\cite{StThSaQuHuMuGr98} we
used Eq.~(\ref{eq:iinelas})  to explain experimental  desorption rates at
negative sample bias due to inelastic hole tunneling. 
Here we present a derivation of Eq.~(\ref{eq:iinelas}),
and use it to calculate desorption rates at positive
sample bias due to inelastic electron tunneling.

\begin{figure}
\leavevmode
\epsfxsize=85mm
\epsffile{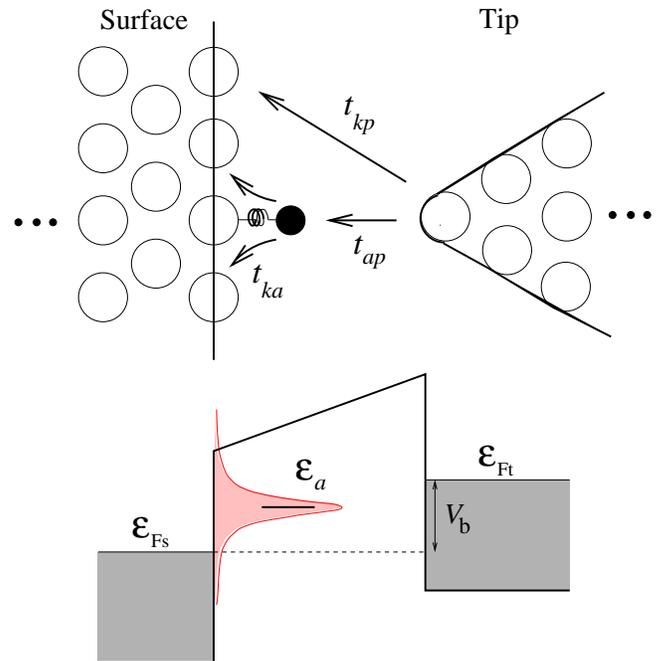}
\caption{Schematic figure of tunneling model (top) and associated 
density of states (bottom).  When the adsorbate (black dot) is separated
from the surface, the eigenstates of the surface and adsorbate are 
$|k\rangle$ and $|a\rangle$, respectively.  Upon adsorption, 
a coupling term with elements $t_{ka}$ is introduced between $|k\rangle$
and $|a\rangle$. In addition, there is a phonon term corresponding to 
the local oscillation of the adsorbate.  The electronic part (i.e., when
there are no phonons) can be diagonalized to yield eigenstates
$|\mu\rangle$. The center of resonance of the broadened level 
$|a\rangle$, $\epsilon_a$, is a function of $z$ the distance of the
adsorbate from its equilibrium position.
} 
\label{fig_schem}
\end{figure}

We divide the tunnel region into a tip and a sample part, 
and assume that the atom to be manipulated 
is adsorbed onto the sample.  The model is shown schematically
in Fig.~\ref{fig_schem}.  The total Hamiltonian is given by
$\hat H = \hat H_{s\!+\!a} + \hat H_{\rm tip} + \hat H_T$, where
the terms on the right are the surface plus adsorbate, tip and 
tunneling parts, respectively.  We first examine $\hat H_{s\!+\!a}$.
Let $|a\rangle$ and $|k\rangle$ be 
the one-electron eigenstates of the {\sl separated} atom and 
the surface, respectively.
When the atom adsorbs onto the surface, 
a coupling term $\hat h_{ka} = \sum_k t_{ka} \hat c_k^\dagger \hat a +
{\rm h.c.}$ is introduced (where $\hat c_k$ and $\hat a$ are the field
operators for states $|k\rangle$ and $|a\rangle$ respectively).  
The surface$+$adsorbate system can be diagonalized to yield new
eigenstates $|\mu\rangle$.  The adsorbate also can vibrate 
on the surface, which implies a localized phonon term 
with a boson field operator $\hat b$.  Furthermore, the adsorbate vibration 
couples to the energy-level of the electron in state $|a\rangle$. 
Therefore, $\hat H_{s+a} = \hat H_e + \hat H_{\rm ph} + \hat H_{e-{\rm ph}}$,
(the electron, phonon and coupling term respectively) where
\begin{mathletters}
\begin{eqnarray}
\hat H_e &=& \sum_\mu \varepsilon_\mu\, \hat c_\mu^\dagger
\hat c_\mu,\\
\hat H_{\rm ph} &=& \hbar \omega_0\, \hat b^\dagger \hat b,\\
\hat H_{e-{\rm ph}} &=& \lambda (\hat b^\dagger + \hat b) \hat a^\dagger 
\hat a,
\end{eqnarray}
\end{mathletters}
The tip Hamiltonian is given by $\hat H_{\rm tip} = \sum_{p} 
\varepsilon_p \hat c_p^\dagger \hat c_p$ (where $|p\rangle$ are
the tip eigenstates) and the tip to surface tunneling by $\hat H_T = 
\sum_{\mu p} t_{\mu p} \hat c_\mu^\dagger \hat c_p + 
{\rm h.c.} \equiv 
\sum_{\kappa p} t_{\kappa p} \hat c_\kappa^\dagger \hat c_p 
+ {\rm h.c.}$ (here, $\kappa = k$ or $a$).
The matrix elements are related via 
$t_{\kappa p} = \sum_\mu \langle \kappa | \mu\rangle t_{\mu p}$.



The coupling between the tip and the surface is normally weak, 
hence it is sufficient to work to lowest (i.e, second) order in the tunneling
matrix element.  Let $|p, n\rangle$ be the state with an electron
in state $p$ on the tip and $n$ phonons on the adsorbate, and 
$|\mu, n'\rangle$ be the {\sl coupled} eigenstate of $H_s$,
arising from the interaction between electron state $\mu$ and 
$n'$ adsorbate phonons.  The inelastic component of the current from tip to 
surface which changes the phonon occupation from $n$ to $n'$ is given
by\cite{Ma90}
\begin{eqnarray}
I(n\rightarrow n') & = & \frac{2 \pi e}{\hbar} \sum_{\mu p}
f_t(\varepsilon_p) (1-f_s(\varepsilon_\mu))\nonumber\\
&&\ \ \ | T_{pn,\mu n'}|^2\delta 
(\varepsilon_p-\varepsilon_\mu-(n'-n)\hbar
\omega_0),
\label{eq:It}
\end{eqnarray}
where $T_{pn,\mu n'}  =   \langle p n | \hat H_T | \mu n'\rangle$
is the $T$-matrix element and
$f_{s,t}(\varepsilon)=1/\{1+\exp[(\varepsilon-\varepsilon_{Fs,Ft})]\}$ are 
Fermi distribution functions. The Fermi levels $\varepsilon_{Fs}$ and
$\varepsilon_{Ft}$ of the sample and tip, respectively,  are determined
by the applied sample bias, $V_{b}$, and related by
$\varepsilon_{Fs}=\varepsilon_{Ft}-eV_{b}$.  

We define uncoupled electron-phonon eigenstates (i.e., setting 
$\lambda = 0$) of the sample as $|\mu \rangle \otimes | n\rangle$. 
The coupled eigenstates can be expressed as a Born series\cite{LiSc50,Sc68} 
\begin{equation}
|\mu, n\rangle 
= \sum_{l=0}^\infty  
\left(\hat \G0(\epsilon_{\mu,n})
\hat H_{e-{\rm ph}}\right)^l |\mu \rangle \otimes | n\rangle ,
\label{eq:bornser}
\end{equation}
where $\epsilon_{\mu,n}\approx \varepsilon_\mu +
n\,\hbar\omega_0$ is the eigenenergy of $|\mu, n\rangle$, and
$\hat \G0 = (\varepsilon - \hat H_e - \hat H_{\rm ph} + i0^+)^{-1}$ 
is the Green function of the uncoupled electron--phonon system. 
Using Eq.~(\ref{eq:bornser}) the $T$-matrix elements can 
be written as
\begin{eqnarray}
\label{eq:T}
T_{pn,\mu n'} 
&=&
\sum_{\mu'} t^*_{\mu' p} \sum_{l = 0}^\infty \langle \mu'|  \otimes
\langle n |  (\hat \G0(\epsilon_{\mu,n}) 
\hat H_{e-{\rm ph}})^l | \mu \rangle \otimes | n'\rangle. 
\label{eq:T1}
\end{eqnarray}

We now consider $n' = n+N$; {\it i.e.,} where $N$ phonons are absorbed
(the emission case is similar).
The first term in the Born series Eq.~(\ref{eq:T1}) which contributes 
is the $N$th order term containing $(\hat b^\dagger)^N$.  In this
paper, we ignore the higher order terms in the series\cite{firstpnote}.
This term gives 
\begin{eqnarray}
&&T_{pn,\mu n+N}^{(N)} = \sqrt{(N+n)!/n!}\ \lambda^N
\ \langle a | \mu \rangle
\sum_{\mu'} t_{\mu' p}\,\times
\nonumber\\ 
&&\langle \mu' |\hat g(\varepsilon_\mu + N\hbar\omega_0)|a\rangle
\!\!\!\prod_{j = 1,N-1}\!\!\! \langle a |
\hat g(\epsilon_{\mu,m}+(N-j)\hbar\omega_0) |a \rangle 
\label{eq:Tapprox}
\end{eqnarray}
where $\hat g(\varepsilon) =  (\varepsilon - \hat H_e +i0^+)^{-1}$
is the sample$+$adsorbate electron Green function (without
phonons).

The Lippmann-Schwinger equation gives the one-electron
eigenstates of the sample$+$adsorbate in terms of the eigenstates of
the atom and the sample,
$|\mu_k \rangle = |k\rangle + \hat g(\varepsilon_{\mu_k})
|a\rangle t^*_{ka}.$  This, together with Eq.~(\ref{eq:Tapprox}) and
the relationship between $t_{\mu p}$ and $t_{\kappa p}$ yields 
\begin{eqnarray}
T^{(N)}_{pn,\mu_k n+N} =&& \sqrt{(n+N)!/n!}\ \lambda^N\, 
[t_{\mu_k p}^* - t_{kp}^*]
\nonumber\\
\times\prod_{j = 0,N-1}&&\!\!
\langle a |\hat g(\varepsilon_{\mu_k} + (N-j)\hbar\omega_0)| a\rangle 
+ {\mit\Upsilon}^{(N)}_{pn,{\mu_k} n+N}.
\label{eq:T(N)}
\end{eqnarray}
The term ${\mit\Upsilon}$ corresponds to processes where the
electron tunnels from the tip into the sample state, $k$,
hops back into the adsorbate,
emits phonons and hops back out again.  Generally, such terms are
small, and moreover it may even be inaccurate to
include this term, since the lifetime of state $k$ probably will be
dominated by scattering with electron-hole excitations or bulk phonon 
states (which is excluded in our model).

The next step is to relate the above equation to
the projectors $\langle a|\mu \rangle$.  We use the
identity
\begin{equation}
|\langle a|\hat g(\varepsilon)|a\rangle|^2 \equiv \frac{\pi\sum_{\mu'} |\langle
\mu'|a\rangle|^2\,\delta(\varepsilon_{\mu'} - \varepsilon)}
{\Delta(\varepsilon)},
\label{eq:identity}
\end{equation}
where $\Delta(\varepsilon)$ is the magnitude of the imaginary part of the self-energy 
of $\langle a |\hat g(\varepsilon)|a\rangle$.
We now assume that the adsorbate resonance is broad compared to the
$\hbar\omega_0$.  Hence we (1) approximate the terms 
$\varepsilon_{\mu_k} + (N-j)\hbar\omega_0 \approx \varepsilon_{\mu_k}$,
(2) assume $\Delta(\varepsilon) \approx \Delta(\tilde\varepsilon_a) \equiv \Delta$,
and (3) set $\sum_{\mu'} |\langle \mu'|a\rangle|^2\,\delta(
\varepsilon_{\mu'} -\varepsilon_{\mu} ) \approx |\langle \mu |a\rangle|^2 \bar\rho_s$  
where $\bar\rho_s = \sum_{\mu'}\delta(\varepsilon-\tilde\varepsilon_a)$
is the density of states at $\tilde\varepsilon_a$. 
These assumptions together with Eqs.~(\ref{eq:It}), (\ref{eq:T(N)}) and
(\ref{eq:identity}) yield
\begin{eqnarray}
I(n\rightarrow n+N) && = \frac{(n+N)!}{n!} \frac{2 \pi e } {\hbar} K^{N} 
\sum_{p\mu_k}  |t_{\mu_k p } - t_{k p}|^2 \times
\nonumber\\
|\langle\mu|a\rangle|^{2 N} &&
f_t(\varepsilon_p)[1-f_s(\varepsilon_{\mu})] 
\delta(\varepsilon_p -\varepsilon_{\mu}-N \hbar \omega),
\label{eq:inn}
\end{eqnarray}
where $K=\pi\bar \rho_s\lambda^2/\Delta$ is a dimensionless quantity.

The last step is to evaluate the matrix element $t_{\mu_k p}-t_{k p}$
using the Tersoff-Hamann approximation\cite{TeHa85,Ch93}. In this 
approximation the tunneling is assumed to be through a single atom at the  
tip apex with an $s$-type wave-function.
We define $|t_{\mu_k p}-t_{k p}|^2 \equiv f_{\mu_k p} |t_{\mu_k
p}|^2$, where $f_{\mu_k} \equiv 
|\langle a|\mu\rangle |^{2}/
(x_{\mu_k}+|\langle a|\mu\rangle |^{2})$ 
is essentially the fraction of electrons
which tunnel from the tip into state $\mu_k$ through the resonance. 
The evaluation of $|t_{\mu_k p}-t_{k p}|^2$ then is
equivalent to previous work on elastic tunneling\cite{StQuGr97} and
for a W tip we obtain
\begin{eqnarray}
I(n \rightarrow n+N) \approx &&  C_{\mathrm{w}} \frac{(n+N)!}{n!} 
\int_{N \hbar \omega_0}^{e V_{\rm
b}} |e^{2R_{\mathrm{w}}\kappa(\varepsilon)}|\, \times\nonumber\\
&&\rho_{N}(d+R_{\mathrm{w}},\varepsilon,E) d\varepsilon.
\label{eq:iinelas}
\end{eqnarray}
In this equation $d$ is the tip-sample distance, 
$\kappa(\varepsilon) = \sqrt{2 m (\phi_t +eV_{\rm b} -\varepsilon)}/\hbar$ 
is the
wave function inverse decay length, $\phi_t=4.5$~eV the work function of the
W tip\cite{CRC94}, $E$ is the electric field between the tip and sample, 
$R_{\mathrm{w}}=3\,$bohr is the atomic radius of W, and the
normalization constant $C_{\mathrm{w}}=0.007 
R_{\mathrm{w}}^2\,$ Amperes $\times$ bohr 
is obtained from a calculation of a model W tip\cite{StQuGr97}. 
The local weighted DOS,
\[ 
\rho_n({\bf r},\varepsilon,E) = K^n\sum_{\mu} f_{\mu} 
|\langle a|\mu\rangle |^{2n} |\psi_{\mu}({\bf r},E)|^2
\delta(\varepsilon-\varepsilon_{\mu}),
\] 
is the DOS weighted by $n$ powers of the dimensionless coupling $K$ 
and the projection $\langle a|\mu\rangle$ of 
the resonance wave function $\psi_{a}$ onto the eigenstates, 
$\psi_{\mu}$, of the sample.
For high biases ($|V_{\rm b}| > 2$ V) the effect of the 
electric field between the tip and the sample, $E$, 
 must be taken into account when calculating
$\psi_{\mu}$\cite{StQuGr97}. The inelastic current
with  energy transfer $n\hbar \omega_0$ is weighted by
$(|\langle a|\mu \rangle |^2 K)^n$, thus only 
eigenstates with a significant overlap with the
adsorbate resonance contribute to the inelastic current.  

We now use Eq.~(\ref{eq:iinelas}) to model the 
STM induced desorption  of H from Si(100)-H(2$\times$1). It has previously been
proposed\cite{ShWaAbTuLyAvWa95} that inelastic scattering of the
tunneling electrons with the  Si-H 6$\sigma$*
resonance  is  the dominant desorption mechanism for sample biases in
the range $2 \,  \mathrm{V} < V_{\rm b} < 4 \,{\rm V}$.  In the following
we  make a first principles calculation of the desorption rate due to
this mechanism as a function
of the tunnel current and sample bias and
compare with experimental  desorption rates. 

The electronic
structure calculations are based on  density-functional theory\cite{HoKo64,KoSh65}  within
the generalized-gradient approximation(GGA)\cite{PeWa91} using 20~Ry plane-wave
basis sets.  We  describe the
Si(100)-H(2$\times$1) surface  by a 12 layer ($2\times1$) slab and
use 32 $k$-points in the  surface Brillouin zone.
Ultra-soft pseudo-potentials\cite{Va90} are used for both H and Si.
The geometry of the 
surface is obtained by relaxing the H atoms and the first six  Si layers,
and the resulting bond lengths and bond angles of the surface atoms 
are similar to other first principles calculations\cite{KrHaNo95,RaCa96}.
From a frozen phonon calculation we obtain the Si-H stretch frequency
$\hbar \omega_0=0.26$~eV, and subtracting   the 
ground state energy of a free H atom we calculate the desorption barrier
$E_{\rm des}=3.36$~eV. 

\begin{figure}
\begin{center}
\leavevmode
\epsfxsize=85mm
\epsffile{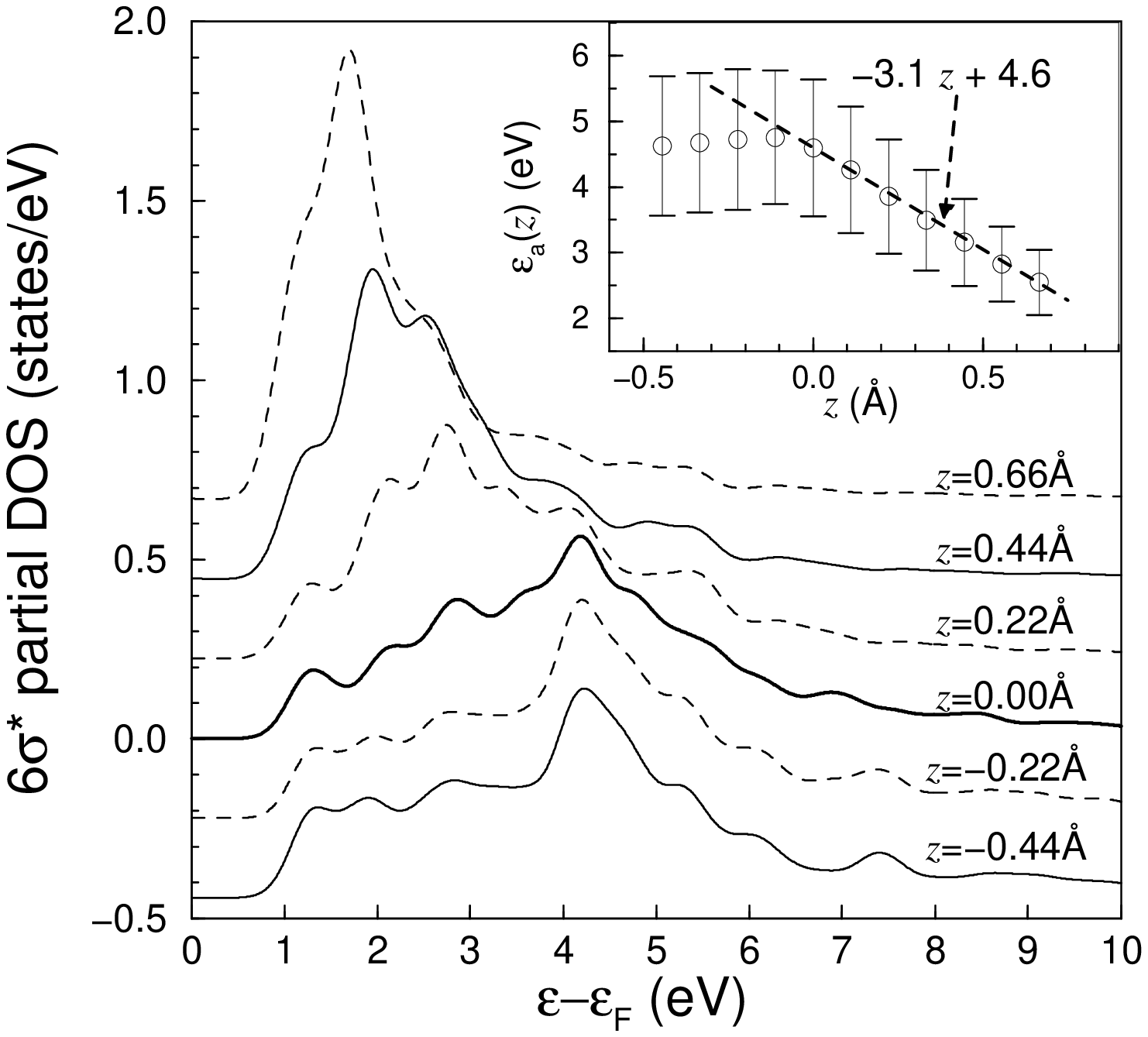}
\end{center}
\caption{ Projections of the Si-H molecular 6$\sigma^*$ eigenstate onto the
  eigenstates, $\mu$, of a Si(100)-H(2$\times$1) slab for different  Si-H
  separations($z=0$ is the equilibrium separation). Only projections of
  states with $\epsilon_{\mu} > \epsilon_{\mathrm{F}}$ are included, and each
 partial DOS is normalized to 1. The inset shows the average centers(circle)
  and widths(error bar) as  determined from  first
  and second moments. The dashed line shows the
  results of a linear least
  squares fit to centers, $\epsilon_a$,  with $z>-0.1$\AA . }
\label{fig_pdos}
\end{figure}

We will use the Si-H molecular $6\sigma*$ eigenstate,
$|m_{6\sigma*}\rangle$, to represent the  $6\sigma*$ resonance.
We expect $|m_{6\sigma*}\rangle$ to have a
large overlap with the  $6\sigma*$ resonance, and smaller
overlaps with the other resonances localized around Si-H, which  are mainly the
Si-H $4\sigma$ and $5\sigma$ resonances. The 4$\sigma$ and
5$\sigma$ resonances are well
below the Fermi level, and we remove these components by 
projecting out the overlap between $|m_{6\sigma*}\rangle$ and
occupied slab eigenstates. 
In Fig.~\ref{fig_pdos} we
show the resulting $6\sigma$* partial DOS for different Si-H
separations, $z$.  From the first and second
moments we determine the average centers, $\tilde{\varepsilon}_a(z)$, and
widths, $\Delta(z)$, as shown in the inset of Fig.~\ref{fig_pdos}. 
At the equilibrium separation, $z=0$,  the resonance is centered at 
$\tilde{\varepsilon}_a=4.6$ eV, and has an average  width $\Delta=1$
eV. For $z>-0.1$ \AA , the variation of
$\tilde{\varepsilon}_a$ with $z$ is nearly linear and from a
linear  least
squares fit in this region we calculate
$\partial {\varepsilon}_a/\partial z =-3.12\pm0.06$ eV/\AA . 
The electron--phonon coupling term  is given by
$\lambda =\sqrt{\hbar/(2M\omega_0)}\partial {\varepsilon}_a/\partial z$,
where $M$ is the adsorbate mass and from the above value of $\partial
{\varepsilon}_a/\partial z$ we obtain $\lambda=0.26$ eV.

To find $\bar \rho_s$ we use the relation
$n_a(\varepsilon) \approx |\langle a| \mu \rangle|^2 \bar \rho_s$.
Considering only the eigenstates with the largest projections
$\langle a| \mu \rangle$, we find  $\bar \rho_s\approx1.5 $ eV$^{-1}$.
To estimate an average value of  $x_{\mu}$, 
we will use the relation $|t_{\mu p}|^2=(|\langle
\mu|a\rangle|^2+x_{\mu})|t_{a p}|^2$. We first select the state,
$\beta$, with largest overlap, $\langle a|\beta\rangle$, in the energy
range of interest. If we  assume $|t_{a p}|^2$ and $x_{\mu}$
are constant in the energy range we obtain the equation
$x_{\mu} = \frac{|\langle\beta|a\rangle|^2 |t_{\mu p}|^2-
|\langle\mu|a\rangle|^2 |t_{\beta p}|^2}{|t_{\beta p}|^2-|t_{\mu
p}|^2},$
which we use to estimate $x$ for each eigenstate in the
energy range $2$V--$4$V.  We find  the average value $x\approx
0.01$ $(\sim 0.1 |\langle a|\beta\rangle|^2)$. With this value we
obtain that $I_0/I \approx 0.5$ for biases in the range $2$V--$4$V;
thus $50\%$ of the elastic electrons passes through the
resonance.

We  now have all the parameters entering
Eq.~(\ref{eq:iinelas}) for the inelastic current.  From
 the inelastic currents we  calculate 
the desorption rate, $R$, by solving
the Pauli master equation for the transitions among the various levels
of the oscillator\cite{GaPeLu92}, and  vibrational relaxations
due to phonon-phonon coupling is described by a current independent
relaxation rate, $\gamma=1 \times
10^8$~s$^{-1}$\cite{ShWaAbTuLyAvWa95,GuLiMi95}. We assume that 
desorption  occurs  when the energy of the
H atom exceeds the desorption energy
$E_{\rm des}=3.36$~eV, corresponding to a truncated
harmonic potential well  with 13 levels.
Our objective is to obtain $R$  
as function of $I$ and
$V_{\mathrm{b}}$. However,  the inelastic current is a function
of $V_{\mathrm{b}}$ and $d$, and we therefore calculate  the
elastic current, 
$I(V_{\mathrm{b}},d)$, to establish a relation between $I$,
$V_{\mathrm{b}}$ and $d$. 
For a given  value of $d$ and $V_{\mathrm{b}}$ we then calculate the inelastic
currents, $I(n\rightarrow n+N)$ and $I(n+N\rightarrow n)$ for all
bound vibrational states $n=1,2,\cdots 13$. We  include events with
$N=1,2,3$, and note  that the $N=1$ term gives the main 
contribution to the desorption rate in the current range relevant for
desorption. 

\begin{figure}
\begin{center}
\leavevmode
\epsfxsize=85mm
\epsffile{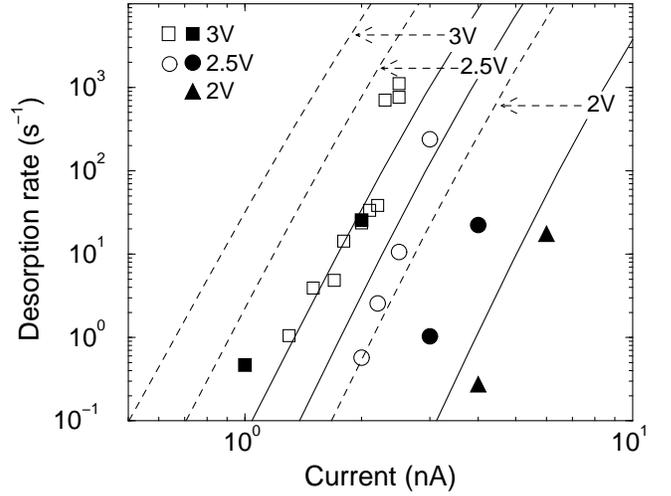}
\end{center}
\caption{ Desorption rate, $R$, as function of current for
  $V_{b}=3.0$~V(square), $V_{b}=2.5$~V(circles) and
  $2.5$~V(triangles). Open  symbols show our experimental data, and 
filled symbols show data from
  Ref.~\protect\cite{ShWaAbTuLyAvWa95}.  Dashed lines
  show  theoretical calculations with first principles value
  $\lambda=0.26$ eV for the electron phonon coupling, while  solid
  lines show results using a fitted value $\lambda=0.20$ eV. }
\label{fig_desrate}
\end{figure}

In Fig.~\ref{fig_desrate} we compare
the calculated  current dependent desorption
rate for sample biases of 2 V, 2.5 V, and  3 V with 
experimental data obtained by us (open symbols) and Shen {\it et al.}
\cite{ShWaAbTuLyAvWa95} (filled symbols). The two experimental data
sets have been obtained on  highly
doped $n$-type samples using similar experimental procedures.
The dashed lines show the first principles calculation of the
desorption rates.  Both experimental and theoretical desorption rates 
follow a power law $R\propto I^\alpha$
with $\alpha \sim 9-10 $ for the theoretical curves and $\alpha \sim
10-13 $ for the experimental data.  However, 
the theoretical curves are shifted towards lower currents.
The discrepancy is due to the three main approximations of the theoretical
model: an
expansion  based on $\Delta \gg \lambda$,$\omega_0$, 
a harmonic approximation for the Si-H bond potential, and  
neglect of excited state orbital relaxation
when calculating the resonance wave function.
In order to obtain theoretical desorption rates comparable to
experiment we   adjust the electron-phonon coupling. Solid lines
in Fig.~3 shows results using  $\lambda=0.20$~eV, and we see that with
this moderate change of the coupling constant we can obtain 
quantitative agreement with the experimental data in the voltage range
2--3 V. Above 3 V, the calculated desorption rates are too small to
explain  experimental desorption rates, and  the measured desorption rates
must be related to another mechanism, most likely 
direct excitation of the Si-H bond as suggested in 
Ref.~\cite{ShWaAbTuLyAvWa95}. 

In conclusion, we have presented a novel first principles theory
of inelastic scattering, and used it to calculate 
the voltage and current dependent variation of the hydrogen desorption
rate from the Si(100)-H(2$\times1$) surface at positive bias
conditions.  We find that the desorption in the voltage range
$2$--$3$ V is consistent with  vibrational heating of the H atom due to  inelastic-scattering
with  the Si-H $6\sigma^*$  resonance.

This work was supported by  the Japanese Science and
Technology Agency, and the use of Danish national computer
resources was supported by the Danish Research Councils.
B.~Y.-K.~Hu and X.~C.~Xie acknowledge a travel grant from 
NATO.

\end{multicols}  

\begin{thebibliography}{10}
\bibitem{firstpnote}
The higher order terms are of order $|T^{N+m}/T^{N}|^2
  =O((\lambda/\Delta)^{2m})$ and can usually be neglected. If needed, these
  terms can also be relatively easily calculated.
\end{thebibliography}
\end{document}